\documentclass[singlespacing]{elsart}

\usepackage{graphicx}

\usepackage{amssymb}
\journal{Computational and Theoretical Chemistry}
\begin{document}

\begin{frontmatter}

\title{Analytic calculation of ground state splitting in symmetric double well potential}

\author{ A.E. Sitnitsky},
\ead{sitnitsky@kibb.knc.ru}

\address{Kazan Institute of Biochemistry and Biophysics, FRC Kazan Scientific Center of RAS,
P.O.B. 30, 420111, Russian Federation. Tel. 8-843-231-90-37. e-mail: sitnitsky@kibb.knc.ru }

\begin{abstract}
The exact solution of the one-dimensional Schr\"odinger equation with symmetric trigonometric double-well potential (DWP) is obtained via angular oblate spheroidal function. The results of stringent analytic calculation for the ground state splitting of ring-puckering vibration in the 1,3-dioxole (as an example of the case when the ground state tunneling doublet is well below the potential barrier top) and 2,3-dihydrofuran (as an example of the case when the ground state tunneling doublet is close to the potential barrier top) are compared with several variants of approximate semiclassical (WKB) ones. This enables us to verify the accuracy of various WKB formulas suggested in the literature: 1. ordinary WKB, i.e., the formula from the Landau and Lifshitz textbook; 2. Garg's formula; 3. instanton approach. We show that for the former case all three variants of WKB provide good accuracy while for the latter one they are very inaccurate. The results obtained provide a new theoretical tool for describing relevant experimental data on IR spectroscopy of ring-puckering vibrations.
\end{abstract}

\end{frontmatter}

\section{Introduction}
Quantum particle transfer in a double-well potential (DWP) is one of the main processes in reaction rate theory. Proton transfer in hydrogen bonds is a notable example of the above general case. The latter takes place in the most important biological molecules such as proteins (participating in some enzymatic reactions \cite{All09}, \cite{Pu06}) and DNA (arguably participating in the occurrence of mutations \cite{Law66}, \cite{God15}). For proton transfer (let alone electron transfer) the effects of quantum tunneling are of utmost importance \cite{Bel80}, \cite{Gol86}, \cite{Ben93}, \cite{Ben94}. The modern approach to taking into account dissipative effects at quantum tunneling is based on the Lindblad master equation for the time evolution of the density matrix and Caldeira-Leggett model of the thermal bath (see \cite{Wei08}, \cite{Han90}, \cite{Ank07}, \cite{Bre02} and refs. therein). Within the framework of proton transfer description this scheme is developed in \cite{Mey87}, \cite{Mey90} and used in \cite{God15}, \cite{God14}, \cite{Sch96}. In it the Hamiltonian of a system under consideration includes the terms
\[
H=...+V(x)+x\sum_{m} f_mq_m+...
\]
where  $V(x)$ is a DWP describing the quantum particle movement along the reaction coordinate $x$ and the latter is linearly coupled to the harmonic oscillators $q_m$ of the thermal bath with $f_m$ being coupling constants. As a result the transition matrix $<i\mid x \mid n>$ becomes one of the key values of the theory. In fact the latter makes use of some transformed matrix element which diagonalises $x$ between the pairs of states that are tunnel doublets \cite{Mey87}, \cite{Mey90}, \cite{God15}, \cite{God14}. However prior obtaining the latter one should have the initial matrix element $<i\mid x \mid n>$. Its calculation requires the knowledge of the eigenfunctions $\psi_n(x)$ (i.e., $\mid n>$) for the Schr\"odinger equation (SE) with DWP $V(x)$. The lack of the exact analytic form of the eigenfunctions $\psi_n(x)$ is the main source of tiresome numeric calculations at practical evaluating the reaction rate constant. The aim of the present article is to suggest a convenient DWP for which SE can be exactly solved thus providing a useful tool for further theoretical analysis.

In the above mentioned proton transfer one actually deals with a real particle. Besides there are many problems in physics and chemistry that can be reduced to SE for a fictitious quantum particle moving in some DWP. A well-known example is the inversion of ammonia molecule $NH_3$ \cite{Hun27}, \cite{Ros32}, \cite{Man35}, \cite{Tow55}, \cite{Vol49}, \cite{Her45}, \cite{Hug00}, \cite{Jan14}. More recent applications include  Bose-Einstein condensates, heterostructures, superconducting circuits and semiconducting quantum rings (see \cite{Xie12}, \cite{Dow13}, \cite{Che13}, \cite{Har14}, \cite{Dow16}, \cite{Dow17}, \cite{Tur10}, \cite{Tur16}, \cite{Col17} and refs. therein). We take into consideration only smooth DWP and pass over those with rectangular wells or two single-wells (harmonic, Morse, etc.) sewed together. The latter models are very helpful in revealing the pertinent physics in numerous systems pertaining in particular to semiconducting devices. Nevertheless their accuracy is always questionable. In contrast smooth hyperbolic or trigonometric DWPs (see below) can provide stringent mathematical treatment of the corresponding SE. However for them one has to deal with complex special functions of mathematical physics. SE with the single-well potentials mentioned above could be solved via habitual hypergeometric function. For DWP one by necessity has to resort to less familiar and more complex mathematical objects such as confluent Heun's function (CHF), spheroidal function (SF) or Coulomb (generalized) SF. Fortunately progress in their realization in mathematical software packages such as {\sl {Mathematica}} ot {\sl {Maple}} does their application to various problems to be routine and convenient.

Recently the reduction of SE with DWP to the confluent Heun's equation (CHE) enabled one to obtain quasi-exact (i.e., exact for some particular choice of potential parameters) \cite{Xie12}, \cite{Dow13}, \cite{Che13} and exact (those for an arbitrary set of potential parameters) \cite{Har14}, \cite{Sit17}, \cite{Sit171} solutions. A plenty of potentials for SE are shown to be exactly solvable via CHF \cite{Ish16}. This special function has been well studied by now and tabulated in {\sl {Maple}} \cite{Fiz12}, \cite{Fiz10}, \cite{Sha12}. It was shown in \cite{Sit171} that equivalently the solution of SE with trigonometric DWP can be expressed via Coulomb (generalized) SF \cite{Kom76}. As a result the obtained solution of SE is very convenient for usage. Earlier CHF was used for obtaining the exact solution of the Smoluchowski equation for reorientational motion in Maier-Saupe DWP \cite{Sit15}, \cite{Sit16} that gives the probability distribution function in the form convenient for application to NNR \cite{Sit11}. In the present article we develop similar approach initiated in \cite{Sit17}, \cite{Sit171} for SE with trigonometric DWP. It should be stressed that the reduction of SE with trigonometric DWP to the Coulomb (generalized) SF requires integer $m$ and thus in this form SE with trigonometric DWP belongs to quasi-exact type. However it was shown in \cite{Sit171} that the solution of SE with trigonometric DWP can be equivalently written via the confluent Heun's function (CHF) that deals with the parameter $h=m^2-\frac{1}{4}$ instead of $m$. CHF is determined at any $h$ including those produced by non-integer $m$. Thus there is no condition on the potential parameter in this case. For this reason one can assert that trigonometric DWP is an exactly solvable one.

The other analytic tool for treating SE with DWP is the well-known semiclassical method (WKB approximation). There are various variants of this method: the ordinary one \cite{Lan74}, \cite{Gar00}, \cite{Par97}, \cite{Par98}, \cite{Son08}, \cite{Ras12}, \cite{Son15}, the numerical realization \cite{Jel12} and the instanton approach \cite{Col85}, \cite{Kle95}, \cite{Tur10}, \cite{Tur16}, \cite{Gil77}, \cite{Neu78}. By now all these variants are considerably matured but still remain an intensive field of investigations. The doubtless merit of WKB is its universal character. Its formulas can be applied to any pertinent DWP. Nevertheless WKB is an approximate method and its accuracy is always under question. One should verify the results either by numerical calculations or by comparison with exact solution if the latter is available. The trigonometric DWP suggested in \cite{Sit17}, \cite{Sit171} is amenable to exact analytic treatment and thus enables one a possibility to verify the accuracy of various WKB formulas. Thus it is useful to verify the results of WKB approach by comparison with those of exact solution of SE with trigonometric DWP. The latter (see \cite{Sit17}, \cite{Sit171}) is a particular variant of a general DWP from \cite{Ish16} (N2 with $m_{1,2}=\left(1/2,1/2\right)$ from Table.1). For trigonometric DWP the wave function is expressed with the help of confluent Heun's function (CHF) or equivalently via Coulomb (generalized) SF. For the symmetric case the latter is actually the angular oblate SF that is implemented in {\sl {Mathematica}} and consequently is very convenient for usage. Besides the spectrum of its eigenvalues is implemented in {\sl {Mathematica}} that enables one to calculate of energy levels for trigonometric DWP very easily \cite{Sit171}. For the asymmetric case the usage of CHF is more helpful at present because Coulomb (generalized) SF is still not realized in a standard package of Mathematica. It is worthy to note that the corresponding package was developed long ago by Falloon \cite{Fal01} but unfortunately up to now is not implemented in the standard package. The latter make its usage to be  a very troublesome problem. If we deal only with the structure of energy levels (for example with the ground state splitting) CHF implementation in {\sl {Maple}} provides convenient usage. However when one tries to calculate the integrals including the wave function with CHF one encounters with a problem \cite{Sit17}, \cite{Sit171} and has to circumvent in a cumbersome way. As a result for the asymmetric case CHF is the only instrument realized in the standard software packages though it should be used with caution \cite{Fiz12}, \cite{Fiz10}. It seems interesting to apply the trigonometric DWP for comparison of the accuracy of various WKB formulas. In particular the case of the tunneling doublet close to the potential barrier top is of interest. WKB is known to be invalid in this case \cite{Lan74} but to quantify the measure of its inaccuracy one needs either the results of numerical solution of SE or that of its exact solution. Trigonometric DWP enables one to obtain exact analytic solution that is in some aspects more convenient for theoretical analysis than its numerical counterpart.

In our earlier articles trigonometric DWP was applied to proton in hydrogen bonds \cite{Sit17} and to ammonia molecule \cite{Sit171}. However in \cite{Sit171} SE was solved in a circumvent two-step process (first the solution was obtained via CHF and then the result was transformed into that via SF). In the present article we find it expedient to provide the direct derivation of the solution via SF that simplifies the presentation considerably. As mentioned above such form of the solution is more convenient for symmetric DWP. Besides such derivation makes the present article to be self-contained. Thus the aim of the article is to provide the exact solution of SE with the trigonometric DWP via SF, to obtain energy levels and their corresponding analytic representation of the wave functions. Then we apply our general results to  ring-puckering vibration in the 1,3-dioxole (experimental data are taken from \cite{Laa09}, \cite{Laa93}) as an example of the case when the ground state tunneling doublet is well below the potential barrier top and to ring-puckering vibration in 2,3-dihydrofuran (experimental data are taken from \cite{Laa09}, \cite{Laa01}) as an example of the case when the ground state tunneling doublet is close to the potential barrier top. For these objects detailed data on IR spectroscopy are available \cite{Laa09}, \cite{Laa93}, \cite{Laa01} that makes them to be good test objects for verifying the accuracy of various theoretical methods. Ring-puckering vibration (see \cite{Laa09}, \cite{Laa93}, \cite{Laa01}, \cite{Laa67}, \cite{Laa671}, and refs.therein) along with ring-twisting, ring-flapping, wagging, torsion and inversion vibrations \cite{Laa09}, \cite{Klo98}, \cite{Oco11},  \cite{Laa87} is an interesting interdisciplinary phenomenon for chemical physics, IR spectroscopy and physics of molecules. It was studied experimentally in details mainly in the laboratory of J. Laane and a lot of experimental data for numerous objects is available at present. Also intensive efforts for its theoretical description have led to the developments in treating the corresponding SE with suitable vibrational potential energy surface that is actually a double minimum function \cite{Laa09} or else DWP. Many workable schemes for numerical solution for SE with DWP have been studied and applied to $2-4$ DWP and its various extensions including in particular the sixth power term \cite{Laa70}.
Also it was shown that in some cases a one-dimensional DWP is insufficient and two-dimensional potential functions are necessary to be considered \cite{Klo98}.
The "2-4" DWP have been thoroughly investigated and applied to numerous objects \cite{Laa09}. It was shown that "2-4" DWP as an example of a two-parameter potential function provides good accuracy in describing pertinent experimental data. Unfortunately SE with "2-4" DWP is not amenable to exact analytic solution. This fact relates the above mentioned field with an old but still poorly solved problem of quantum mechanics to obtain stringent analytic description for the motion of a quantum particle in a DWP. The main aim of the article is to compare the obtained exact calculation for the ground state splitting in the above mentioned two cases with those of several WKB formulas. We make thorough comparison of our exact analytic result with those of ordinary WKB variant (the formula from Landau and Lifshitz textbook for the ground state splitting in the case of a symmetric DWP \cite{Lan74}), Garg's formula \cite{Gar00} and the instanton method \cite{Col85}, \cite{Kle95}, \cite{Gil77}, \cite{Neu78}.

\section{WKB formulas}
In this preliminary Sec. we present several formulas for the WKB estimates of the ground state splitting in symmetric DWP available in the literature. The well known formula for the ground state splitting in the case of a symmetric DWP $V(x)$ is given in \cite{Lan74}
\begin{equation}
\label{eq1} E_1-E_0=\frac{\hbar\omega}{\pi} \exp\left[-\frac{\sqrt {2M}}{\hbar}\int_{-\bar c}^{\bar c}dx\ \left \vert\sqrt{V(x)-E_m}\right \vert\right]
\end{equation}
Here $E_m=\left(E_1-E_0\right)/2$, $\bar c$ is the turning point corresponding to $E_m$ and $\omega$ is the classical vibration frequency for the bottom of the well of the potential $V(x)$.

The author of \cite{Gar00} suggested an very useful formula
\begin{equation}
\label{eq2}  E_1-E_0=\hbar\omega \left(\frac{M\omega \bar d^2}{\pi\hbar}\right)^{1/2}\exp\left(A-S_0/\hbar\right)
\end{equation}
Here $\pm \bar d$ are the locations for the minima of DWP,
\begin{equation}
\label{eq3} S_0=\int_{-\bar d}^{\bar d}dx\ \sqrt{2M\left[V(x)-V\left(\bar d\right)\right]}
\end{equation}
and
\begin{equation}
\label{eq4} A=\int_{0}^{\bar d}dx\ \left[\frac{M\omega}{\sqrt{2M\left[V(x)-V\left(\bar d\right)\right]}}-\frac{1}{\bar d-x}\right]
\end{equation}

The instanton method \cite{Col85} provides for the ground state splitting the formula
\begin{equation}
\label{eq5}  E_1-E_0=2\hbar K\left(\frac{S_0}{2\pi\hbar}\right)^{1/2}\exp\left(-S_0/\hbar\right)
\end{equation}
where $S_0$ is given by the expression (\ref{eq3}). The standard way to obtain the value of
$K$ is \cite{Col85}
\begin{equation}
\label{eq6}  K=\left[\frac{\rm{det}\left(-\partial ^2_\tau+\omega^2\right)}{\rm{det'}\left[-\partial ^2_\tau+M^{-1}V''\left(x_{cl}(\tau)\right)\right]}\right]^{1/2}
\end{equation}
where $det'$ indicates that the zero eigenvalue is to be omitted at computing the determinant. The instanton $x_{cl}$ is obtained from the classical equation of motion
\begin{equation}
\label{eq7} M\frac{d^2x_{cl}(\tau)}{d\tau^2} -V'\left(x_{cl}\right)=0
\end{equation}
that has a solution
\begin{equation}
\label{eq8} \tau=\sqrt {\frac{M}{2}}\int_{0}^{x_{cl}}\frac{dx}{\sqrt{V(x)-V\left(\bar d\right)}}
\end{equation}
It obeys the boundary conditions $x_{cl}\left(-T/2\right)=-\bar d$ and $x_{cl}\left(T/2\right)=\bar d$.
Garg showed that (\ref{eq2}) is equivalent to the formula given by the instanton approach \cite{Col85}. However it was shown within the framework of Coleman's approximation
\begin{equation}
\label{eq9} K\approx \sqrt{2\omega}\beta
\end{equation}
where $\beta$ is a coefficient given by the asymptotic behavior of the instanton velocity at $\tau\rightarrow \infty$
\begin{equation}
\label{eq10} \left(\frac{M}{S_0}\right)^{1/2}\frac{dx_{cl}}{d\tau}\approx \beta \exp \left(\omega \tau \right)
\end{equation}

\section{Solution of Schr\"odinger equation with trigonometric DWP}
We treat the one-dimensional SE for a fictitious quantum particle with the reduced mass $M$
\begin{equation}
\label{eq11} \frac{d^2 \psi (x)}{dx^2}+\frac{2M}{\hbar^2}\left[E-V(x)\right]\psi (x)=0
\end{equation}
where $V(x)$ is a DWP. The latter is assumed to be infinite at the boundaries of some finite interval $x=\pm L$.
Further we use the dimensionless values for the distance $y$, the potential $U(y)$ and the energy $\epsilon$,
\begin{equation}
\label{eq12} y=\frac{\pi x}{2L}\ \ \ \ \ \ \ \ \ \ \ \ \ \ \ \ \ U(y)=\frac{8ML^2}{\hbar^2 \pi^2}V(x)\ \ \ \ \ \ \ \ \ \ \ \ \ \epsilon=\frac{8ML^2E}{\hbar^2 \pi^2}
\end{equation}
where $-\pi/2\leq y \leq \pi/2$. We consider trigonometric DWP
\begin{equation}
\label{eq13} U(y)=h\ \tan^2 y-b\sin^2 y+a\sin y
\end{equation}
As we remain within the finite interval  $-\pi/2\leq y \leq \pi/2$ (physically it means the, e.g., at ring-puckering vibration in 1,3-dioxole or in 2,3-dihydrofuran the covalent bonds between the atoms are not broken and our fictitious quantum particle can not leave the finite interval at the boundaries of which the potential becomes infinite) then we can safely discard the periodic character of trigonometric functions.
For the symmetric case ($a=0$) of the trigonometric DWP the parameters $h$ and $b$ are related with the barrier height $B=-U\left(y_{min}\right)$ and barrier width $\Delta=y_{min}^{(1)}-y_{min}^{(2)}$ as follows
\[
b=\frac{B}{\left\{1-\left[\cos\left(\Delta/2\right)\right]^2\right\}^2} \ \ \ \ \ \ \ \ \ \ \ \ \ \ \ \ \ \ \ \ \ \ \ h=\frac{B\left[\cos\left(\Delta/2\right)\right]^4}{\left\{1-\left[\cos\left(\Delta/2\right)\right]^2\right\}^2}
\]
Inversely we obtain
\[
\Delta=2\arccos\left(\frac{h}{b}\right)^{1/4}\ \ \ \ \ \ \ \ \ \ \ \ \ \ \ \ \ \ \ \ \ \ \ \ B=\left(\sqrt {h}-\sqrt {b}\right)^2
\]
The dimensionless form of SE with trigonometric DWP (\ref{eq13}) is
\begin{equation}
\label{eq14} \psi''_{yy} (y)+\left[\epsilon-h\ \tan^2 y +b\ \sin^2 y-a\ \sin y\right]\psi (y)=0
\end{equation}

We introduce the designations
\begin{equation}
\label{eq15}b=p^2
\end{equation}
\begin{equation}
\label{eq16}h=m^2-\frac{1}{4}
\end{equation}
The examples of the potential (\ref{eq13}) for a symmetric case are depicted in Fig.1 and Fig.2.
\begin{figure}
\begin{center}
\includegraphics* [width=\textwidth] {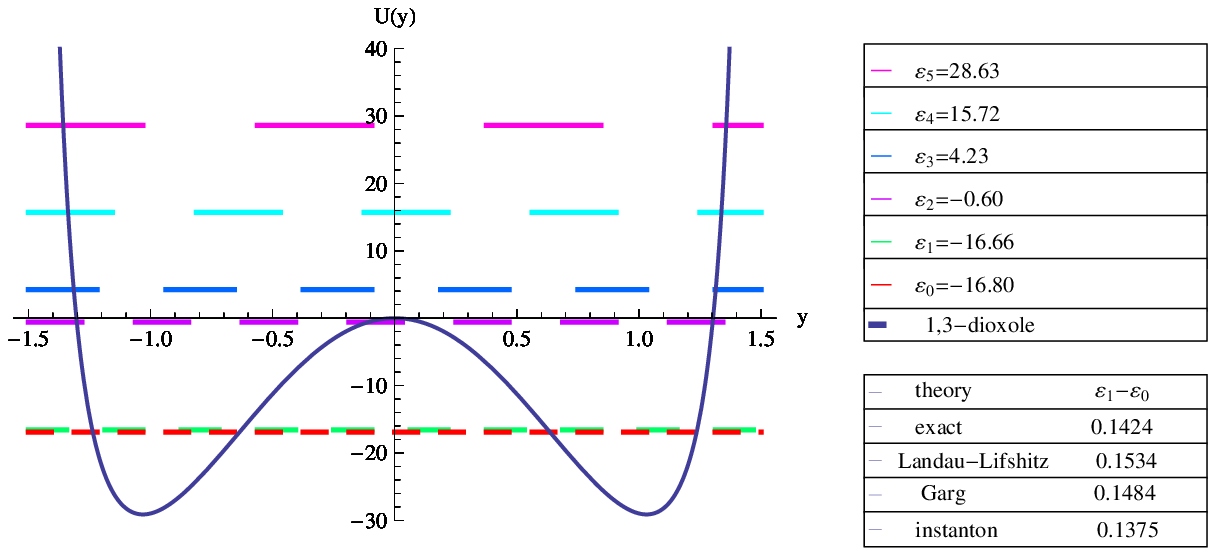}
\end{center}
\caption{The model double-well potential (\ref{eq13}) in designations (\ref{eq15}), (\ref{eq16}) at the values of the parameters $m=2$, ($h=3.75$), b=53.78, ($p\approx 7.33$), $a=0$. The parameters are chosen to describe the potential and the energy levels for ring-puckering vibration in the 1,3-dioxole (experimental data are taken from \cite{Laa09}, \cite{Laa93}) as an example of the case when the ground state tunneling doublet is well below the potential barrier top. The energy levels $\epsilon_0=-16.80$, $\epsilon_1=-16.66$, $\epsilon_2=-0.60$, $\epsilon_3=4.23$, $\epsilon_4=15.72$, $\epsilon_5=28.63$ are respectively depicted by the dashes of increasing length. The  splitting of the ground state $\epsilon_1-\epsilon_0=0.1424$ corresponds to $2.1\ {\rm cm^{-1}}$ in dimensional units. The results of calculations within the framework of different WKB approaches are also indicated: Landau and Lifshitz textbook formula $\epsilon_1-\epsilon_0=0.1534$;  Garg's formula $\epsilon_1-\epsilon_0=0.1484$; instanton approach $\epsilon_1-\epsilon_0=0.1375$. For this case WKB results provide good accuracy.} \label{Fig.1}
\end{figure}
\begin{figure}
\begin{center}
\includegraphics* [width=\textwidth] {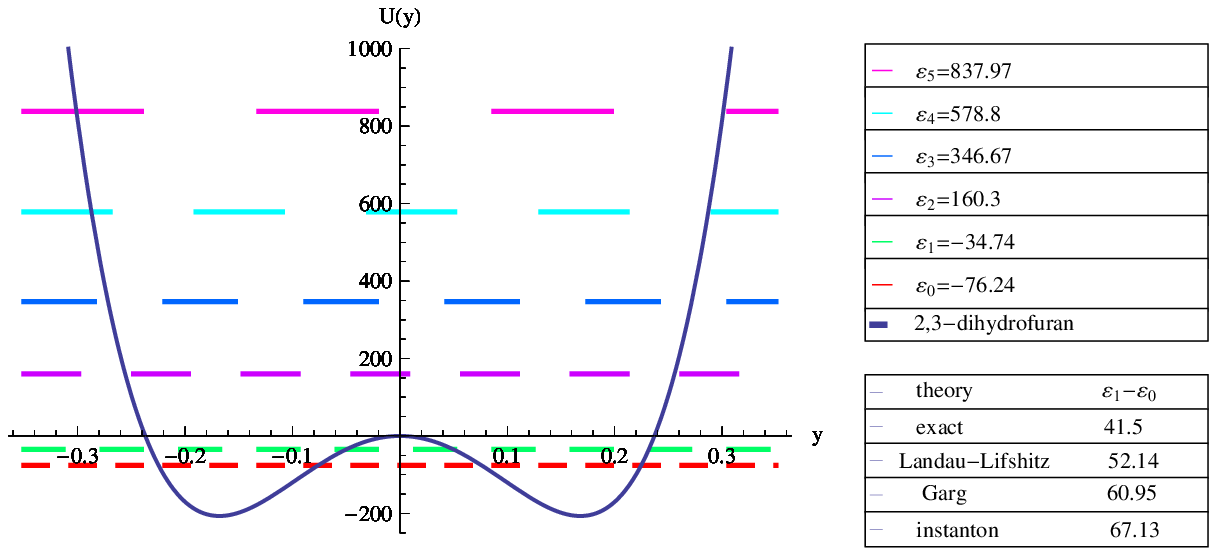}
\end{center}
\caption{The model double-well potential (\ref{eq13}) in designations (\ref{eq15}), (\ref{eq16}) at the values of the parameters $m=500$, ($h=249999.75$), $b = 264583$, ($p\approx 514.37$), $a=0$. The parameters are chosen to describe the potential and the energy levels for ring-puckering vibration in the 2,3-dihydrofuran (experimental data are taken from \cite{Laa09}, \cite{Laa01}) as an example of the case when the ground state tunneling doublet is close to the potential barrier top. The energy levels $\epsilon_0=-76.24$, $\epsilon_1=-34.74$, $\epsilon_2=160.30$, $\epsilon_3=346.67$, $\epsilon_4=578.80$, $\epsilon_5=837.97$ are respectively depicted by the dashes of increasing length. The stringent calculation from the exact analytic solution of the  corresponding Schr\"odinger equation for the ground state  splitting $\epsilon_1-\epsilon_0=41.5$ corresponds to $18.9\ {\rm cm^{-1}}$ in dimensional units. The results of calculations within the framework of different WKB approaches are also indicated: Landau and Lifshitz textbook formula $\epsilon_1-\epsilon_0=52.14$; Garg's formula $\epsilon_1-\epsilon_0=60.95$; instanton approach $\epsilon_1-\epsilon_0=67.13$. For this case WKB results are very inaccurate.} \label{Fig.2}
\end{figure}
We use a new variable
\begin{equation}
\label{eq17} s=\sin y
\end{equation}
where $-1\leq s \leq 1$ and introduce a new function $v(s)$
\begin{equation}
\label{eq18} \psi (y)=v(\sin y)\cos^{1/2} y
\end{equation}
The equation for $v(s)$ takes the form
\begin{equation}
\label{eq19} \frac{d}{ds}\left[\left(1-s^2\right)\frac{dv(s)}{ds}\right]+\Biggl\{
\epsilon+m^2-\frac{1}{2}-as+p^2s^2-
\frac{m^2}{1-s^2}\Biggr\}v(s)=0
\end{equation}

At integer $m$ (\ref{eq19}) is a Coulomb (generalized) spheroidal equation \cite{Kom76}
and its solution can be written as
\begin{equation}
\label{eq20} v(s)=\bar\Xi_{mq}\left(p, -a;s\right)
\end{equation}
Here $q=0,1,2,...$ and $\bar\Xi_{mq}\left(p, -a;s\right)$ is CSF. The energy levels are determined by the relationship
\begin{equation}
\label{eq21} \epsilon_q=\lambda_{mq}\left(p, -a\right)+\frac{1}{2}-m^2-p^2
\end{equation}
Here $\lambda_{mq}\left(p, -a\right)$ is the spectrum of eigenvalues for the function $\bar\Xi_{mq}\left(p, -a;s\right)$ (see Appendix).
The wave function is
\begin{equation}
\label{eq22} \psi_q (y)=\cos^{1/2} y\ \bar\Xi_{mq}\left(p, -a;\sin y\right)
\end{equation}

For the case of symmetric DWP ($a=0$) eq. (\ref{eq19}) is an angular oblate spheroidal equation \cite{Kom76} (one can also consider it as the limit $ \bar \Xi_{mq}\left(p, 0;s\right)=\bar S_{m(q+m)}\left(p;s\right)$ in the above formulas)
and its solution has the form
\begin{equation}
\label{eq23} v(s)=\bar S_{m(q+m)}\left(p;s\right)
\end{equation}
Here $q=0,1,2,...$ and $\bar S_{m(q+m)}\left(p;s\right)$ is the angular oblate SF. The latter is implemented in {\sl {Mathematica}} as $\bar S_{m(q+m)}\left(p;s\right)\equiv \rm{SpheroidalPS}[(q+m),m,ip,s]$. The energy levels are determined by the relationship
\begin{equation}
\label{eq24}
\epsilon_q=\lambda_{m(q+m)}\left(p\right)+\frac{1}{2}-m^2-p^2
\end{equation}
Here $\lambda_{m(q+m)}\left(p\right)$ is the spectrum of eigenvalues for $\bar S_{m(q+m)}\left(p;s\right)$. It is implemented in {\sl {Mathematica}} as $\lambda_{m(q+m)}\left(p\right)\equiv \rm{SpheroidalEigenvalue}[(q+m),m,ip]$.
For the ground state splitting we obtain
\begin{equation}
\label{eq25} \epsilon_1-\epsilon_0=\lambda_{m(1+m)}\left(p\right)-\lambda_{mm}\left(p\right)
\end{equation}
As a result we have
\begin{equation}
\label{eq26} \psi_q (y)=\cos^{1/2} y\ \bar S_{m(q+m)}\left(p;\sin y\right)
\end{equation}

The formulas  (\ref{eq24}), (\ref{eq25}) and (\ref{eq26}) provide a very convenient and efficient tool for calculating the energy levels, the ground state splitting and the wave functions for SE in the case of symmetric trigonometric DWP (\ref{eq13}) with the use of {\sl {Mathematica}}.

It is worthy to note the following circumstance that is highly important for practical applications of the above formulas. Both the functions $\bar\Xi_{mq}\left(p, -a;s\right)$ and $\bar S_{m(q+m)}\left(p;s\right)$ as defined in \cite{Kom76} are normalized by the requirements
\begin{equation}
\label{eq27} \int_{-1}^1ds\  \bar\Xi_{mq'}\left(\sqrt {b}, -a;s\right)\bar\Xi_{mq}\left(\sqrt {b}, -a;s\right)=\delta_{qq'}
\end{equation}
\begin{equation}
\label{eq28} \int_{-1}^1ds\ \bar S_{m(q+m)}^2\left(\sqrt{b};s\right)=1
\end{equation}
Hence from the purely theoretical point of view the wave functions (\ref{eq22}) and (\ref{eq26}) are normalized.
In contrast the realization of the function $S_{m(q+m)}\left(p;s\right)$ in {\sl {Mathematica}} as $SpheroidalPS[(q+m),m,ip,s]$ is not normalized. For this reason at practical application of the the wave function (\ref{eq26}) below we conceive it as a preliminary normalized one
\begin{equation}
\label{eq29}  \psi_i(y)^{normalized}=\psi_i(y)\left\{\int_{-\pi/2}^{\pi/2}dy\ \left(\psi_i(y)\right)^2\right\}^{-1/2}
\end{equation}
\begin{figure}
\begin{center}
\includegraphics* [height=5cm] {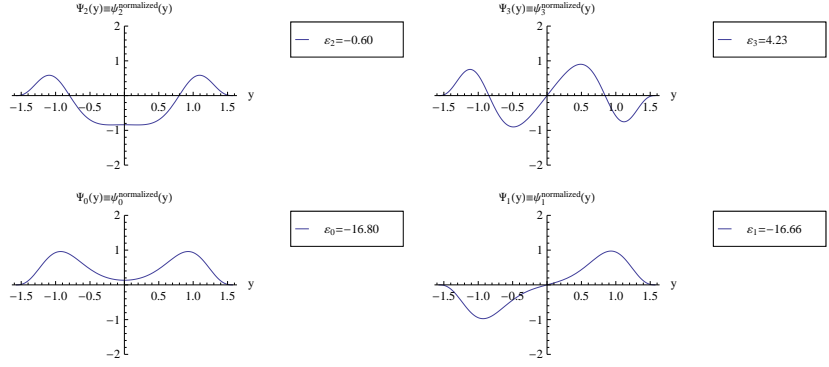}
\end{center}
\caption{Normalized wave functions (\ref{eq26}) for Schr\"odinger equation (\ref{eq14}) with the double-well potential (\ref{eq13}) corresponding to the energy levels $\epsilon_0=-16.80$, $\epsilon_1=-16.66$, $\epsilon_2=-0.60$, $\epsilon_3=4.23$. The parameters of the potential $m=2$, ($h=3.75$), b=53.78, ($p\approx 7.33$), $a=0$ are chosen to describe the energy levels for ring-puckering vibration in the 1,3-dioxole (experimental data are taken from \cite{Laa09}, \cite{Laa93}).} \label{Fig.3}
\end{figure}
The example of the behavior of the wave functions is presented in Fig.3.

\section{WKB formulas in dimensionless form}
For the comparison of the accuracy of WKB formulas with our exact result (\ref{eq25}) it is convenient to cast them into dimensionless form.
We rewrite the formula from Landau and Lifshitz textbook (\ref{eq1}) in the dimensionless form with the help of (\ref{eq2}). To obtain the relationship of the frequency $\omega$ in the vicinity of the minimum $x_{min}$ with the derivative of the dimensionless potential $U(y)$ at $y=y_{min}$ we expand it into a series
\begin{equation}
\label{eq30} V(x)
\left| {\begin{array}{l}
  \\
{x\sim x_{min}}\\
 \end{array}}\right. \approx V\left(x_{min}\right)+\frac{M\omega^2}{2}\left(x-x_{min}\right)^2
\end{equation}
Then
\begin{equation}
\label{eq31} \omega=\frac{\hbar \pi^2}{4ML^2}\sqrt{\frac{1}{2}\left(\frac{d^2U(y)}{dy^2}\right)\left| {\begin{array}{l}
  \\
{y_{min}}\\
 \end{array}}\right. }
\end{equation}
We denote the dimensionless analog of $\bar c$ as $c$
\begin{equation}
\label{eq32}  c=\frac{\pi \bar c}{2L}
\end{equation}
Then the  ground state splitting in the dimensionless from is
\begin{equation}
\label{eq33} \epsilon_1-\epsilon_0=\frac{2}{\pi}\left(\frac{1}{2}\left (\frac{d^2 U(y)}{dy^2}\right)\left| {\begin{array}{l}
  \\
{y_{min}}\\
 \end{array}}\right. \right)^{1/2}\exp\left(-\int_{c}^{c}dy\ \left \vert \sqrt{U(y)-\epsilon_m}\right\vert\right)
\end{equation}

We rewrite the Garg's formula(\ref{eq2}) in the dimensionless form. The expression for the frequency $\omega$ is given by (\ref{eq31}).
We denote the dimensionless analog of $\bar d$ as $d$
\begin{equation}
\label{eq34} d=\frac{\pi \bar d}{2L}
\end{equation}
Then the ground state splitting in the dimensionless form is
\[
\epsilon_1-\epsilon_0=2d\left(\frac{1}{2}\left (\frac{d^2 U(y)}{dy^2}\right)\left| {\begin{array}{l}
  \\
{y_{min}=d}\\
 \end{array}}\right. \right)^{3/4}\times
\]
\[
\exp\Biggl\{\int_{0}^{d}dy\ \Biggl[\left(\frac{1}{2}\left (\frac{d^2 U(y)}{dy^2}\right)\left| {\begin{array}{l}
  \\
{y_{min}=d}\\
 \end{array}}\right. \right)^{1/2}\times
\]
\begin{equation}
\label{eq35} \frac{1}{\sqrt{U(y)-U(d)}}-\frac{1}{d -y}\Biggr]-\int_{-d}^{d}dy\ \sqrt{U(y)-U(d)}\Biggr\}
\end{equation}

We rewrite the  formula of the instanton approach (\ref{eq5}) in the dimensionless form. The determinants in the formula (\ref{eq6}) can be  calculated as the products of the eigenvalues of a corresponding equation. This standard way can not be used in the case of trigonometric DWP. The reason is that in contrast to, e.g., "2-4" potential the corresponding equation for (\ref{eq13}) can not be solved. Fortunately for trigonometric DWP the eigenfunctions of SE are known and given by (\ref{eq26}). As a result we have an alternative and direct way for obtaining  $K$. From \cite{Col85} we know the formulas
\begin{equation}
\label{eq36}  <-\bar d \mid \exp \left(-HT/\hbar \right) \mid -\bar d> =\sqrt{ \frac{\omega}{\pi \hbar}}\ e^{-\omega T/2}
 \rm{ch}\left[KT\exp \left(-S_0/\hbar \right)\right]
\end{equation}
\begin{equation}
\label{eq37}  <\bar d \mid \exp \left(-HT/\hbar \right) \mid -\bar d> =\sqrt{ \frac{\omega}{\pi \hbar}}\ e^{-\omega T/2}
 \rm{sh}\left[KT\exp \left(-S_0/\hbar \right)\right]
\end{equation}
Here $H$ is the hamiltonian for SE (\ref{eq1}). Making use of these formulas we can express $K$ as follows
\begin{equation}
\label{eq38} K=\frac{1}{T}\exp \left(S_0/\hbar \right)\ \rm{arcth}\left(\frac{<\bar d \mid \exp \left(-HT/\hbar \right) \mid -\bar d>}{ <-\bar d \mid \exp \left(-HT/\hbar \right) \mid -\bar d>}\right)
\end{equation}
Substituting (\ref{eq38}) into (\ref{eq5}) we obtain
\begin{equation}
\label{eq39}  E_1-E_0=\frac{2\hbar}{T}\left(\frac{S_0}{2\pi\hbar}\right)^{1/2}\ \rm{arcth}\left(\frac{<\bar d \mid \exp \left(-HT/\hbar \right) \mid -\bar d>}{ <-\bar d \mid \exp \left(-HT/\hbar \right) \mid -\bar d>}\right)
\end{equation}
Our eigenfunctions $\mid q>\equiv\psi_q (x)$ given by (\ref{eq26}) make use of the dimensionless coordinate $y$. The latter is related to the dimensional $x$ by (\ref{eq12}). Taking it into account we have
\begin{equation}
\label{eq40} \Biggl <-\bar d \left \vert \exp \left(-\frac{HT}{\hbar} \right) \right \vert -\bar d \Biggr> =\sum_{n=0}^{\infty}\ e^{-\frac{\left(E_n-V\left(\bar d\right)\right) T}{\hbar} }  <-\bar d \mid n><n \mid -\bar d>
\end{equation}
\begin{equation}
\label{eq41}  \Biggl <\bar d \left \vert \exp \left(-\frac{HT}{\hbar} \right) \right \vert -\bar d \Biggr> =\sum_{n=0}^{\infty}\ e^{-\frac{\left(E_n-V\left(\bar d\right)\right) T}{\hbar} } <\bar d \mid n><n \mid -\bar d>
\end{equation}
The wave functions in the locations of the potential minima are $\mid \bar d>=\delta \left(x-\bar d \right)$ and $\mid -\bar d>=\delta \left(x+\bar d \right)$ respectively.

For further progress we need to know $T$. We will use dimensionless values from now on. We introduce the dimensionless time as
\begin{equation}
\label{eq42}  s=\tau\frac{\pi^2\hbar}{4ML^2}
\end{equation}
and rewrite (\ref{eq8}) in dimensionless units
\begin{equation}
\label{eq43} s=\int_{0}^{y_{cl}}\frac{dy}{\sqrt{U(y)-U\left( d\right)}}
\end{equation}
We introduce the dimensionless analog of $T$ by the requirement $y_{cl}(S/2)=d$ and obtain the following relationship for $S$
\begin{equation}
\label{eq44} S=2\int_{0}^{d}\frac{dy}{\sqrt{U(y)-U\left( d\right)}}
\end{equation}
As a result we obtain
\begin{equation}
\label{eq45}  \Biggl <-\bar d \left \vert \exp \left(-\frac{HT}{\hbar} \right) \right \vert -\bar d \Biggr> =\sum_{n=0}^{\infty}\ e^{-\left(\epsilon_n-U\left( d\right)\right) S/2 } \left[\psi_n (-d)\right]^2
\end{equation}
\begin{equation}
\label{eq46}  \Biggl <\bar d \left \vert \exp \left(-\frac{HT}{\hbar} \right) \right \vert -\bar d \Biggr> =\sum_{n=0}^{\infty}\ e^{-\left(\epsilon_n-U\left( d\right)\right) S/2 } \psi_n (d)\psi_n (-d)
\end{equation}
Substitution of all results in (\ref{eq39}) yields
\[
\epsilon_1-\epsilon_0=\frac{4}{S}\left(\frac{1}{2\pi}\int_{-d}^{d}dy\ \sqrt{U(y)-U(d)}\right)^{1/2}\times
\]
\begin{equation}
\label{eq47} \rm{arcth} \left\{\frac{\sum_{n=0}^{\infty}\ e^{-\left(\epsilon_n-U\left( d\right)\right) S/2 } \psi_n (d)\psi_n (-d)}{\sum_{n=0}^{\infty}\ e^{-\left(\epsilon_n-U\left( d\right)\right) S/2 } \left[\psi_n (-d)\right]^2}\right\}
\end{equation}
In (\ref{eq47}) one should use the normalized wave functions obtained from (\ref{eq26}) by the standard procedure (\ref{eq29}).

\section{Discussion and conclusions}
Fig.1 and Fig.2 show that the parameters of the potential (\ref{eq3}) can be chosen to provide good description of the energy levels structure for a set of specific experimental data. In Fig.1 the energy levels for ring-puckering vibration in the 1,3-dioxole (experimental data are taken from \cite{Laa09}, \cite{Laa93}) as an example of the case when the ground state tunneling doublet is well below the potential barrier top are indicated. The transition frequencies are $E_3-E_0=208.6\ {\rm cm^{-1}}$, $E_3-E_2=47.9\ {\rm cm^{-1}}$ and $E_2-E_1=158.6\ {\rm cm^{-1}}$. The ground state splitting is $E_1-E_0=2.1\ {\rm cm^{-1}}$. These experimental values are obtained from our dimensionless ones if we take $m=2$ ($h=3.75$), $b=53.78$ ($p\approx 7.33$), $a=0$. The exact result of stringent analytic solution of SE is $\epsilon_1-\epsilon_0=0.1424$. For 1,3-dioxole $c=0.63$ (see Fig.1). The calculation of (\ref{eq21}) (Landau and Lifshitz textbook formula) yields
$\epsilon_1-\epsilon_0=0.1534$. For 1,3-dioxole $d=1.03$ (see Fig.1). The calculation of (\ref{eq26}) (Garg's formula) yields $\epsilon_1-\epsilon_0=0.1484$.
Taking into account in (\ref{eq42}) (instanton approach) only six lowest energy levels (i.e., approximating $\infty$ by $5$ in the sums) we obtain for 1,3-dioxole
$\epsilon_1-\epsilon_0=0.1375$.

One can see that for the case when the ground state tunneling doublet is well below the potential barrier top all three variants of WKB approach provide good accuracy. However it should be stressed that the formula from the Landau and Lifshitz textbook (ordinary WKB method) formula works well only if the necessary input information is available (the turning points corresponding to $E_m=\left(E_1-E_0\right)/2$). These turning points a priory are unknown and their obtaining poses an additional problem. Thus the formula from the Landau and Lifshitz textbook (ordinary WKB method) requires preliminary calculations of the turning points corresponding to $E_m=\left(E_1-E_0\right)/2$ that creates some inconvenience at its usage. Garg's formula \cite{Gar00} for a symmetric potential (\ref{eq2}) yields slightly better estimate of the ground state splitting than that (\ref{eq1}) from the Landau and Lifshitz textbook  \cite{Lan74}. Besides (\ref{eq2}) indeed has a considerable advantage (noted in \cite{Gar00}) compared with (\ref{eq1}). In the latter the integration is carried out between the turning points corresponding to $E_m$. In contrast in (\ref{eq2}) the integration is carried out between the minima the potential that are known from the shape of DWP. We conclude that the Garg's formula \cite{Gar00} both provides both good accuracy and is very convenient for usage.

Instanton approach \cite{Col85} even in the most elaborate cases ("$2-4$" potential \cite{Col85}, \cite{Gil77} or the "pendulum $1-cos$" one \cite{Neu78}) produces very cumbersome formulas that are difficult for application. For our trigonometric DWP such calculations are impossible but the knowledge of the exact solution of SE enables us to circumvent the difficulties in an alternative way. The result is comparably accurate with that of the Garg's formula and the calculations are rather tedious.
Within the framework of Coleman's approximation (\ref{eq9}) Garg showed that (\ref{eq2}) is equivalent to the formula (\ref{eq5}) given by the instanton approach \cite{Col85}. On our approach we do not use this approximation. As a result we obtain that the estimate based on the Garg's formula differs from that of instanton approach. We conclude that Garg's formula provides the same accuracy as the instanton approach but is much more convenient for applications.

In Fig.2 the energy levels for ring-puckering vibration in the 2,3-dihydrofuran (experimental data are taken from \cite{Laa09}, \cite{Laa01}) as an example of the case when the ground state tunneling doublet is close to the potential barrier top  are indicated. The transition frequencies are $E_2-E_1=88.8\ {\rm cm^{-1}}$ and $E_3-E_0=191.2\ {\rm cm^{-1}}$. The ground state splitting is $E_1-E_0=18.9\ {\rm cm^{-1}}$. These experimental values are obtained from our dimensionless ones if we take
$m=500$, ($h=249999.75$), $b = 264583$, ($p=514.37$), $a=0$. The exact result of stringent analytic solution of SE is  $\epsilon_1-\epsilon_0=41.5$. For 2,3-dihydrofuran $c=0.064$ (see Fig.2). The calculation of (\ref{eq21}) (Landau and Lifshitz textbook formula) yields $\epsilon_1-\epsilon_0=52.14$. For 2,3-dihydrofuran $d=0.168$ (see Fig.2). The calculation of (\ref{eq26}) (Garg's formula) yields $\epsilon_1-\epsilon_0=60.95$. Taking into account in (\ref{eq42}) (instanton approach) only six lowest energy levels (i.e., approximating $\infty$ by $5$ in the sums) we obtain for 2,3-dihydrofuran $\epsilon_1-\epsilon_0=67.13$. One can see that for the case when the ground state tunneling doublet is close to the potential barrier top the WKB approach is very inaccurate.

One can conclude that the Schr\"odinger equation with symmetric trigonometric double-well potential is exactly solved via angular oblate spheroidal function. Our stringent analytic description of the ground state splitting enables one to verify the accuracy of several WKB formulas available in the literature. The exact solution suits well for the description of ring-puckering vibrations as is exemplified by 1,3-dioxole and 2,3-dihydrofuran. Thus it yields a new theoretical tool for interpreting relevant experimental data on IR spectroscopy of such molecules.

\section{Appendix}
\begin{figure}
\begin{center}
\includegraphics* [width=\textwidth] {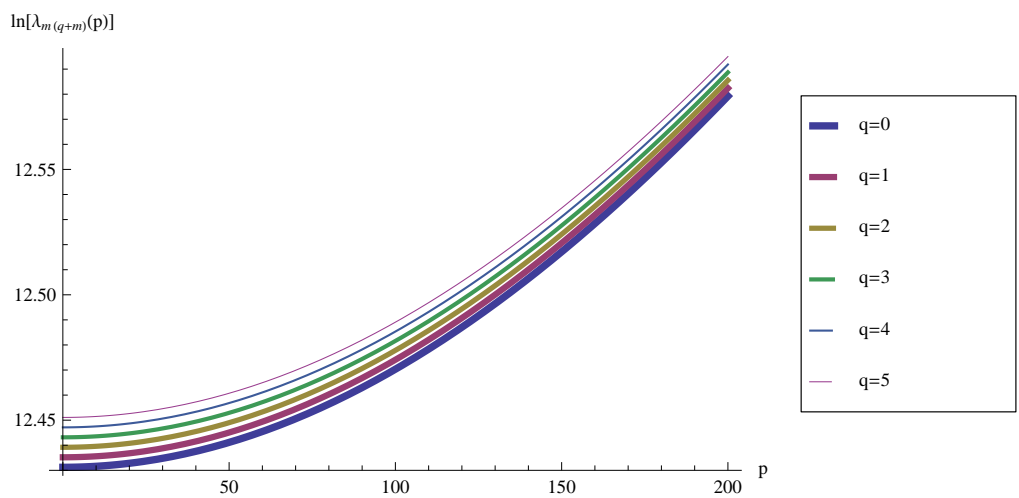}
\end{center}
\caption{The behavior of the discrete spectrum of eigenvalues $\lambda_{m(q+m)}\left(p\right)$ for the spheroidal function $\bar S_{m(q+m)}\left(p;s\right)$ \cite{Kom76} on the parameter $p$ for six energy levels from the lower one $q=0$ (thick line) to the upper one $q=5$ (thin line). The value of the parameter $m$ is $m=500$ as in our example for ring-puckering vibration in the 2,3-dihydrofuran.} \label{Fig.4}
\end{figure}
In this Appendix we briefly present some relevant information on $\lambda_{mq}\left(p, -a\right)$ and $\lambda_{m(q+m)}\left(p\right)$ from \cite{Kom76} and \cite{Fal01}. The quantity $\lambda_{mq}\left(p, -a\right)$ is the infinite discrete spectrum of eigenvalues for the Sturm-Liouville problem that is set for the Coulomb (generalized) spheroidal equation (19) for the corresponding Coulomb (generalized) spheroidal function $\bar\Xi_{mq}\left(p, -a;s\right)$ \cite{Kom76}. There is no explicit form of the closed transcendental equation for $\lambda_{mq}\left(p, -a\right)$ either in\cite{Kom76} or in \cite{Fal01}. Instead of that there are equations of the continued fraction type for various particular cases. Also numerous plots for $\lambda_{mq}\left(p, -a\right)$ as functions of parameters are presented there and the analytic forms for limiting cases are discussed in details \cite{Kom76}, \cite{Fal01}. In the case of symmetric potential $a=0$ (that is the subject of the present article) $\lambda_{mq}\left(p, -a\right)$ is reduced to $\lambda_{m(q+m)}\left(p\right)$. The latter is implemented in {\sl {Mathematica}} as $\lambda_{m(q+m)}\left(p\right)\equiv \rm{SpheroidalEigenvalue}[(q+m),m,ip]$. In this case the asymptotic expansion of $\lambda_{m(q+m)}\left(p\right)$ for large barrier heights $p>>1$ is given by a very cumbersome expression (5.75) from \cite{Kom76}. However it should be stressed that \cite{Kom76} had been written before the realization $ \rm{SpheroidalEigenvalue}[(q+m),m,ip]$ in {\sl {Mathematica}} became available. At present it is much more convenient to use at practical calculations the latter option than the explicit form of the asymptotic expansion. The example of the plot for $\lambda_{m(q+m)}\left(p\right)$ at a given value of the parameter $m$ as a function of the parameter $p$ is presented in Fig.4.

Acknowledgements. The author is grateful to Dr. Yu.F. Zuev
for helpful discussions. The work was supported within the framework of State Assignment (Project N 0217-2018-0009).

\end{document}